# Superconducting exchange coupling driven bistable and absolute switching

Sonam Bhakat[1] and Avradeep Pal*[1,2]

[1] Department of Metallurgical Engineering and Materials Science, Indian Institute of Technology Bombay, Powai, Mumbai, Maharashtra – 400076, India

[2] Centre of Excellence in Quantum Information, Computation, Science and Technology, Indian Institute of Technology Bombay, Powai, Mumbai, Maharashtra – 400076, India

**Abstract:** As per de Gennes' predictions, a superconducting layer placed between two ferromagnetic insulators can drive an antiferromagnetic exchange coupling between them. Using two ferromagnetic insulating GdN layers having dissimilar switching fields sandwiching a superconducting Vanadium thin film, we demonstrate evidence of such exchange coupling. We demonstrate that such an exchange coupling promotes switching between zero and finite resistance states of Vanadium. Our devices hold either a finite resistance or a zero-resistance state at zero magnetic field, dependent on their magnetic field history. Moreover, we demonstrate the absolute switching effect, thus making such devices suitable for application at the lowest temperatures as non-volatile cryogenic memory useful for futuristic quantum circuits and for several other superconducting spintronic applications.

**Introduction**

Spin dependent magnetoresistance (MR) effects in spacer layer separated magnetic multilayers has led to several novel phenomena and device concepts[1–4]. Such devices attract great interest both from a fundamental as well as an applied standpoint. While from a fundamental perspective - novel magnetic exchange coupling mechanisms have been unearthed[5]; the development of such spintronic devices has ushered in a new era in memory technologies. The archetypal device in the current in-plane (CIP) geometry is a Giant Magnetoresistance pseudo spin valve (CIP GMR PSV)[6] where the MR of the device is dependent on the relative orientation between metallic spacer layer separated ferromagnets. Curiously, theoretical predictions of a superconducting spacer layer mediated exchange coupling between two ferromagnetic insulating (FI) layers[7] potentially leading to infinite magneto-resistance, predates much of the experimental work on metallic spintronics related to GMR. However, despite the early predictions, there is very little experimental literature on the subject. With the advent of newer low temperature computing paradigms, it is envisaged that dissipation less cryogenic memory[8] will be a key element for future low temperature computing control architectures, and hence there is a need to comprehensively explore such device proposals in greater details and explore more and better materials options to achieve the same.

The specific proposal under consideration in this work is that of de Gennes[7]. The salient propositions from his theory are i) That there should be an angular dependence of $T_c$ on the relative magnetization orientations of FI layers in an FI/S/FI trilayer superconducting pseudo spin valves (SPSVs) ii) The angular dependence arises due to a relative angular magnetic orientation dependent effective exchange field in the S layer iii) In case such $T_c$ differences exist, the two FI layers can be exchange coupled through the S layer, and its strength is inversely proportional to the S spacer layer thickness. The above physics is aptly encapsulated in the following relation:

$$\bar{h} = 2|\Gamma|S \frac{a}{d_s}\cos\frac{\theta}{2} \tag{1}$$

Equation 1 gives the value of the average exchange field in FI/S/FI trilayers; where $\Gamma$ is the exchange integral, $d_s$ is the thickness of superconducting spacer layer, $\theta$ is the angle between the magnetization of the two F layers, a is the lattice parameter of the superconductor, and S is the ferromagnetic spin.

Among the three propositions, the first two have been verified experimentally for the parallel (P) and anti-parallel (AP) relative magnetic orientations in $Fe_3O_4/In/Fe_3O_4$[9], $EuS/Al/EuS$ trilayers[10] and potentially in GdN/NbN/GdN devices with a Gd interlayer[11]. However, till date, a comprehensive experimental proof of all the three propositions, including that of Superconducting Exchange Coupling (SEC) has only ever been explored in one publication using a GdN/Nb/GdN system[12].

---

*Author to whom correspondence should be addressed: avradeep@iitb.ac.in

In this communication, we show that GdN/V/GdN SPSVs can be used to verify all of de Gennes' propositions; and in the CIP geometry – these SPSVs can operate as a non-volatile cryogenic memory element. Most importantly, we demonstrate that on reduction of V layer thickness, we approach the regime of the absolute spin valve - where the device exhibits a superconducting transition only for the AP state of the ferromagnets, and no superconducting transition is evident for the P state.

A ferromagnetic insulator proximity coupled to a superconductor is a reasonably well understood system[13–16], and the proximity effect in these systems is known to be manifested by an effective Zeeman field inside the superconductor[17]. Among FI material candidates showing evidence of induced Zeeman fields, there is now a considerable body of work on various Europium chalcogenides[10,17–24], and GdN[25,26]. The choice of V as the superconductor is instigated by the need to verify SEC in an alternative superconductor other than Nb. Additionally, V with bulk $T_c$ of 5.4K and coherence length of approximately 44nm, offers a reasonable range of temperatures and experimentally realizable SPSV thicknesses. Moreover, V has been used earlier for designing superconducting spin valves with metallic ferromagnets[27].

**Results and discussion**

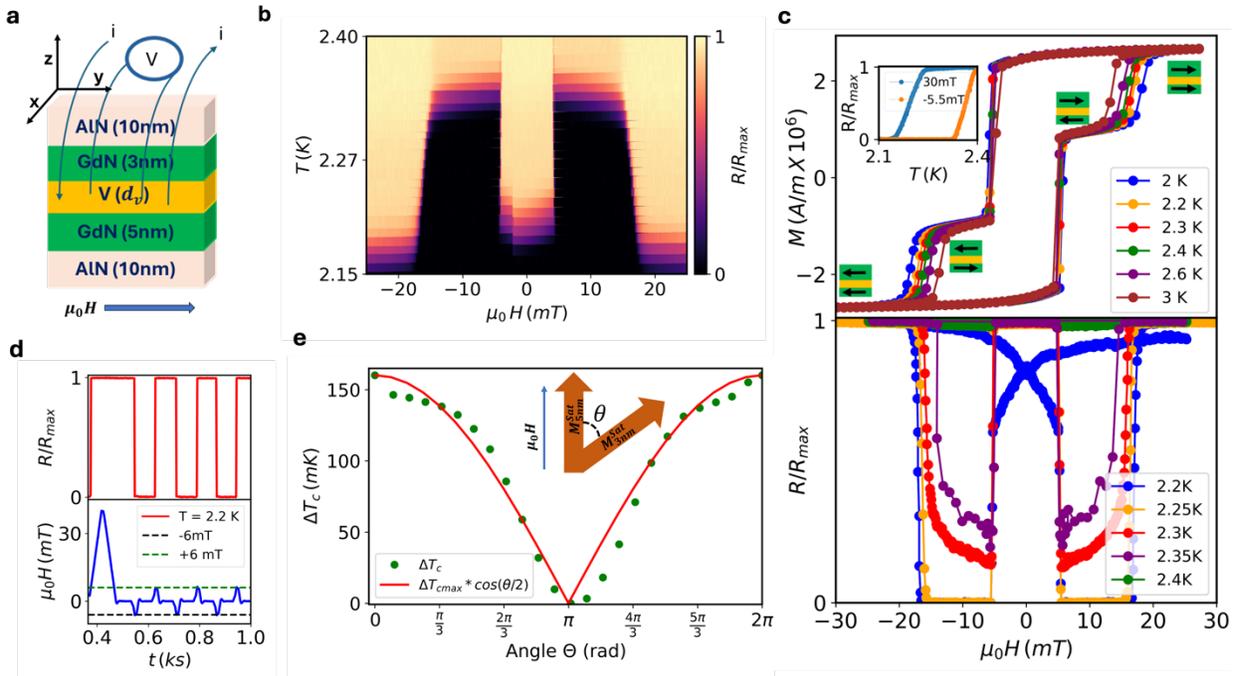

Figure 1: a) A cartoon depiction of a typical SPSV multi-layered stack wire-bonded for four probe measurements b) Colour map of 11nm V thickness sample, where several Resistance (R) vs in plane external magnetic field ($\mu_0 H$) sweeps in the sequence +25mT to -25mT and from -25mT to +25mT (termed downward and upward sweeps respectively) were performed with fixed temperatures; at 5mK intervals from 2.15K to 2.4K. The color map shows the parts of MR measurements corresponding only to 0 to -25mT, and 0 to +25mT, from the downward and upward sweeps respectively c) Top panel shows MH loops of the 11nm SPSV, measured at various temperatures. Inset to top panel of c shows resistance (R) vs temperature (T) measured when the relative orientation of the magnets are in P (+25mT) and AP (-5.5mT) configuration. Lower panel shows R vs H sweeps at various temperatures d) Demonstration of nonvolatile

zero field bistable states in the 11nm SPSV e) Measured value of $\Delta T_c$ of the 11nm SPSV as a function of angle between saturation magnetization vectors of the two GdN layers, plotted along with an ideal $\cos{\theta/2}$ functional behavior of $\Delta T_c$.

In Figure 1 we show the characteristics of a typical GdN/V/GdN superconducting PSV valve with V thickness as 11nm. Figure 1b shows the evolution of resistance (and hence $T_c$) of the system as the in plane magnetic field is swept from zero to negative (positive) values on the left (right) halves of the plot. The magnetic fields are swept after applying a positive (negative) saturation field of 40mT. The color contrast boundary indicates that there is a sudden rise of $T_c$ at approximately 5.5mT on either side, and a relatively more gradual fall of $T_c$ which stabilizes approximately at 16mT on either side. The maximum and minimum $T_c$ of the trialyer is approximately 2.35K and 2.17K respectively. In Figure 1c, we show the MH loops of the same trilayer at various temperatures from 2K to 3K (top panel), along with the MR measurement of the same trilayer from temperatures just above the AP state $T_c$ (2.4K) to temperatures close to P state $T_c$ (2.2K). There are several noticeable features in Figure 1c. First, the drop in magnetization at -5.5mT is much larger as compared to -16mT. This confirms the assumption that 5nm GdN has a lower coercive field than 3nm GdN. Second, there are two drops in magnetizations starting at approximately -5.5mT and -16mT. The first drop in terms of magnetic field almost exactly corresponds to the sudden rise of $T_c$ in Figure 1b (indicating a robust stability of superconducting state on achieving an AP state). Third - these drops in magnetization correspond almost exactly to switching in and out of the superconducting state in the RH measurements. Fourth, in the temperature range between P and AP state $T_c$, the switching field of the 5nm GdN remains almost constant, while that of the 3nm GdN keeps on increasing monotonically with lowering of temperature. This indicates that the onset of superconductivity in the system is correlated with a more prolonged survival of the effective AP state between the two GdN layers. This is a strong indicator for the existence of superconducting exchange coupling (SEC) in the system. To confirm the existence of SEC, we perform $T_c$ measurements on this device as a function of the relative saturation magnetization orientations. In accordance to Equation 1, and because $\Delta T_c \propto \bar{h}$, we expect the same functional form for $\Delta T_c$. In Figure 1e, we show the measured $\Delta T_c$; which tally almost exactly to an expected $\cos{\theta/2}$ dependence.

We note that in Figure 1c, the transitions of the V layer into and out of the superconducting state are remarkably sharp and hence would be ideal for switching applications. Following this observation, in Figure 1d, we show the switching characteristics of the 11nm V trilayer over several cycles. The lower panel shows the sequence of external in plane magnetic field application. The measurement starts with saturating both GdN layers at 40mT, when the V layer remains in a metallic state due to P orientation of both GdN layers. Then several sequences of switching are carried out between – (+) 6mT, thereby repeatedly achieving AP and P state of the trilayer. After application of each field, we return to zero field and record the resistance state of the device. Clear sharp transitions into and out of the superconducting state are observed at – (+) 6mT, and interestingly the same state is retained at zero fields. This clearly demonstrates the suitability of the trilayer as a non-volatile cryogenic memory.

In Figure 2, we show details of measurement of spin valves of several V thicknesses, and elucidate on further evidence of SEC in our system. The crux of SEC is that the onset of superconductivity in the system mediates an effective antiferromagnetic (AF) exchange interaction between the two FI layers. While the onset of superconductivity is ensured by achieving an AP state by the switching of the softer FI layer (5nm GdN); from thereon – the superconducting state (with thickness much lower than the bulk coherence length) couples the two FI layers through coherent electrons of the cooper pair, each sitting at the top and bottom S/FI interfaces. A further experimental evidence of SEC induced AF exchange from MH and RH measurements would be that due to this new AF exchange coupling, it becomes harder than normal to come out of the AP state, and hence a higher-than-normal switching field is required to switch the 3nm layer, in order to break out of the AP state. Through MH and RH measurements in Figure 1c, this phenomenon is already demonstrated for the 11nm V sample. In Figure 2 we demonstrate this in all other measured spin valves. Since it is difficult to access MH below 2K in most commercial magnetometers, and since the switching in RH and MH match for a sample for whom both measurements could be done for a reasonable temperature range; we base most of our observations on switching fields derived from RH measurements in color plots similar to that shown in Figure 1b for all other SPSVs. All color plots of other SPSVs are shown in the supplementary material. A clear increase in switching field of the 3nm layer is visible in the respective $\Delta T_c$ regions; indicating that it indeed becomes harder to come out of the superconducting (AP) state. Interestingly there seems to be almost no noticeable change in the first switching fields of 5nm GdN in all SPSVs in the $\Delta T_c$ region. These observations are almost identical to those in the GdN/Nb/GdN system[12].

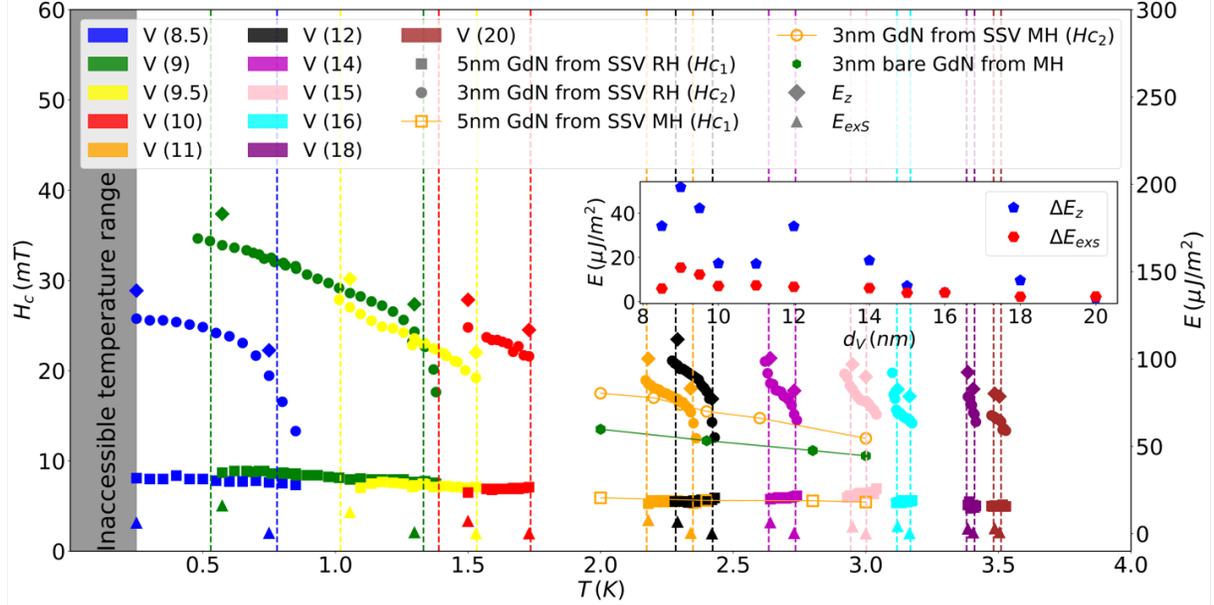

Figure 2: Dashed vertical lines denote the region of operation of SPSVs of various V thicknesses. All quantities of each V thickness are represented by a particular color but different shapes. Filled circles indicate switching field of 3nm GdN from RH measurements. Unfilled circles indicate switching fields of 3nm GdN from MH measurements. Filled squares indicate switching fields of 5nm GdN from RH measurements. Unfilled squares indicate switching fields of 5nm GdN from MH measurements (MH data is only available for the 11nm sample). Filled diamonds indicate switching energy ($E_Z$) associated with switching of the 3nm GdN in each spin valve at the P and AP state Tc. Filled triangles indicate the superconducting condensation energy ($E_{exS}$) associated with the P and AP states of each spin valve. Filled hexagons indicate the switching field of a 3nm GdN sample grown on AlN and capped by AlN.

Phenomenologically, this effect can be understood from a thermodynamical viewpoint. The onset of superconductivity in the AP state introduces the superconducting condensation energy in the system ($E_{exS}$). For mediating the net AF exchange coupling, the increased energy for switching ($E_Z$) of the 3nm GdN out of the AP state to the P state must be compensated by $E_{exS}$. Hence, in the temperature range of operation of SPSVs; $\Delta E_{exS}(T_{cAP} - T_{cP})$ should be comparable to the magnitude of $\Delta E_Z(T_{cAP} - T_{cP})$. Using the expressions for the energy terms per unit area:

$$\Delta E_{exS}(T_{cAP} - T_{cP}) \cong 1.5\gamma d_V T_{cAP}(T_{cAP} - T_{cP}) \tag{3}$$

Where $\gamma$ is the specific heat constant of $V^{28} = 9.8 mJ mol^{-1} K^{-2}$, and all other terms are defined earlier.

$$\Delta E_Z(T_{cAP} - T_{cP}) = 2d_{GdN}M_{GdN}\mu_0(H_{c2-P} - H_{c2-AP}) \tag{4}$$

Where, $d_{GdN}$ = 3nm, $M_{GdN}$ = 0.9*10$^6$ A/m, $H_{c2-P}$ =18.55 mT corresponding to switching field for 3nm GdN at 2.18K, and $H_{c2-AP}$ = 15.39 mT corresponding to switching field of 3nm GdN at 2.34K.

Using the above values, we find that $\Delta E_{exS}(T_{cAP} - T_{cP}) = 7.31\mu Jm^{-2}$, while $\Delta E_Z(T_{cAP} - T_{cP}) = 17.07\mu Jm^{-2}$. The closeness of these two energy scales is similar to that in GdN/Nb/GdN systems[12], and gives a phenomenological justification for the possibility of SEC in our system for the 11nm SPSV. The inset to Figure 2 shows these two energy scales plotted for all SPSVs. We note that for most of the samples these two energy scales are closely matched. For some samples especially between 8.5 and 10 nm, although there seems to be an apparent mismatch, we note that the two energy scales are roughly of the same order of magnitude.

We point out the crucial role of $\gamma$ of the superconducting layer in observation of SEC, as it significantly impacts the magnitude of $\Delta E_{exS}$. Among the elemental superconductors, V has one of the highest magnitudes of $\gamma$, and is significantly higher than that of Nb – the superconductor used for the first experimental demonstration of SEC. This was the main basis for choosing V for this experiment. It is therefore understandable why FI based SPSVs explored earlier with superconductors having almost an order of magnitude lower specific heats – Al[10] ($\gamma$ = 1.35 $mJmol^{-1}K^{-2}$) and In[9] ($\gamma$ = 1.67 $mJmol^{-1}K^{-2}$) may not show typical SEC induced AP state hardening features as shown in Figure 2. On similar lines, we predict that among elemental superconductors, Ta with $\gamma$ = 6.15 $mJmol^{-1}K^{-2}$ and relatively high coherence length of 95nm should be a strong contender as another alternative superconductor for observation of SEC.

Finally, we turn our attention to the possibility of realization of absolute switching using these SPSVs. The original proposal pertaining to absolute spin valve effect in CPP geometry relies on varying magnitudes of tunneling currents between two ferromagnet proximitized superconducting/metal layers separated by a tunnel barrier[29]. In such a device, depending on the relative orientation of the ferromagnets, a finite tunneling current appears or ideally disappears completely at the induced minigap voltage. The CIP equivalent of such a device would be the appearance or complete disappearance of the superconducting state, which is dependent on the relative orientation of ferromagnets. Such a condition can be realized in our SPSVs if the induced exchange fields in the V layer far exceeds its superconducting gap value and well beyond the paramagnetic limit of superconductivity[30,31] for V.

As indicated in previous works by Hauser[9], and Li[10], the P state $T_c$ of the spin valves can be used to estimate Γ and subsequently $\bar{h}$ for each V thickness. For estimating Γ we use:

$$\frac{T_{cP}}{T_{c0}} = 1 - 10\left(\frac{\Gamma S}{E_F}\right)\left(\frac{\xi_{eff}}{d_V}\right) \tag{5}$$

Where $T_{cP}$ is the P state $T_c$ for a particular Vanadium thickness ($d_V$) spin valve, $T_{c0}$ is the corresponding $T_c$ of a bare V film of thickness $d_V$; $E_F$ is the fermi energy of V (approximately 10eV) estimated using free electron theory; $\xi_{eff}$ is the zero-temperature dirty limit coherence length of each bare V film estimated from perpendicular critical field measurements shown in the supplementary information section. Figure 3a shows Γ calculated using the above equation for various V thicknesses. Using the value obtained for Γ, we use equation 1 to estimate the exchange field ($\bar{h}$) in each spin valve. While Γ fluctuates in the range of 150-250meV, $\bar{h}$ rises monotonically by more than an order of magnitude for the lowest thickness. From this observation, we expect the lowest thicknesses of V spin valves to demonstrate absolute switching. In Figures 3b and 3c, we show evidence of such absolute switching in the 8.5nm V SPSV. No evidence whatsoever of appearance of a superconducting state is visible at the lowest temperatures. Although we have a range of inaccessible temperatures (0.26K); we point to the fact that $\Delta T_c$ rises monotonically with reducing thickness (inset to Figure 3a), and hence the $\Delta T_c$ of the 8.5nm SPSV should be greater than the 9nm SPSV (0.8K). This implies that since $T_{cAP}$ of 8.5nm SPSV is 0.78K, and because $\Delta T_c$ is expected to be greater than 0.8K, therefore the $T_{cP}$ should not be observable even till 0K. Finally, in Figure 3d, we demonstrate bistable switching at 0.26K and zero field, with several different patterns of field cycling. This clearly demonstrates the suitability of application of such SPSVs for usage as sub kelvin non-volatile memory elements.

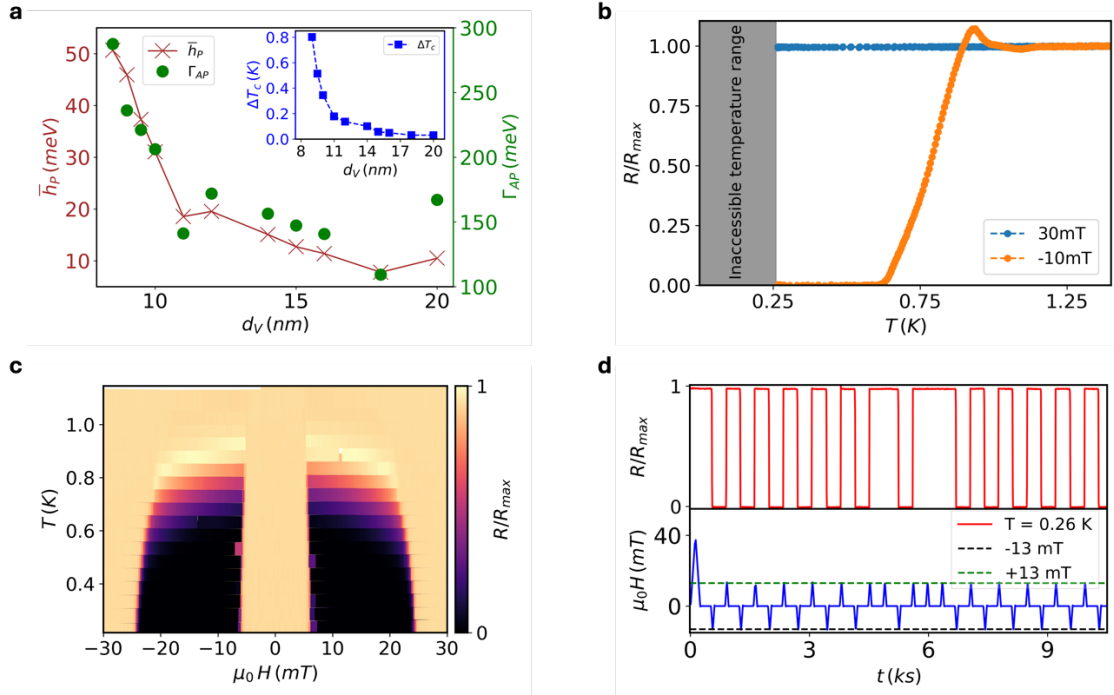

Figure3: a) Calculated value of exchange fields (maroon crosses) in different thicknesses of V layer SPSVs (left axes). Right axes show calculated values of the exchange constant (green circles) for every spin valve as per de Gennes formulation. Inset shows monotonic evolution of $\Delta T_c$ with lowering of V thickness down to 9nm b) RvsT heating measurement of the 8.5nm SPSV at in plane external field values corresponding to P and AP states c) Color plot with RH measurements as a function of temperature for the 8.5nm SPSV d) Demonstration of zero field bistable switching phenomena at 0.26K.

In conclusion, we have demonstrated that V based SPSVs tally with all of de Gennes' predictions, and it mediates SEC between FI layers. Apart from demonstration of a new material system for observation of SEC, the highlight of this work lies in the demonstration of absolute switching and switchable bistable states at the lowest temperatures. Although evidence of bistable switching in SPSVs was shown earlier in an EuS/Al/EuS SPSV[27]; we believe that SEC mediated AF exchange coupled SPSVs makes the non-volatile bistable states more robust and insulate them against typical field cycling effects related to domain wall dynamics and minor loop magnetization changes[32]. Moreover, switching in these SPSVs is remarkably sharp and hence well suited for practical applications as cryogenic memory devices.

**Methods**
Several multilayers of the type AlN/GdN (5nm)/V/GdN(3nm)/AlN with varying V (8nm to 20nm) layer thickness were grown on n-doped Si substrates with a 285nm thermal oxide. The film growth was carried out in an ultra-high vacuum custom designed sputtering system with four DC magnetrons and one RF magnetron, with a base pressure below $2*10^{-9}$ mbar. All samples were grown in the same sputtering run without breaking vacuum. The bottom and top AlN layers act as buffer layer for growth of GdN and capping layer to prevent atmospheric degradation of GdN respectively. While AlN was grown using RF sputtering from an AlN target, GdN was grown using a 92%Ar and 8% N2 reactive gas mixture using a Gd target, and V was grown using Ar gas plasma from a V target. Following the only experimental work on SEC demonstration, GdN thicknesses were kept as 5nm and 3nm to achieve a PSV kind of behavior with 5nm GdN expected to have a lower coercive field as compared to 3nm GdN. All low-temperature resistance measurements were done in Oxford Teslatron PT cryostat, where a base temperature of 0.26K can be achieved using a He3 insert. Un-patterned multilayers were wire bonded and measured in a 4-probe geometry using source measure units with a constant current of $100\mu A$ and the output voltage was recorded. Magnetization measurements were performed in a Quantum Design SQUID magnetometer system.

**Acknowledgements**
This work was financially supported by a Core Research Grant from the Department of Science and Technology, Science and Engineering Research Board, India (Grant No. CRG/2019/004758).

**Author contributions**
SB and AP designed the experiments. SB grew the thin film multilayers and performed all magneto-transport measurements. SB and AP together analyzed the data and wrote the manuscript.

**Data availability**
Data available upon request from corresponding author.

**Competing interests**
The authors declare no competing interests.

**Supplementary Information to:**
**Superconducting exchange coupling driven bistable and absolute switching**
Sonam Bhakat[1] and Avradeep Pal*[1,2]

[1] Department of Metallurgical Engineering and Materials Science, Indian Institute of Technology Bombay, Powai, Mumbai, Maharashtra – 400076, India
[2] Centre of Excellence in Quantum Information, Computation, Science and Technology, Indian Institute of Technology Bombay, Powai, Mumbai, Maharashtra – 400076, India

1. **Colour maps of SPSVs of various thicknesses:**

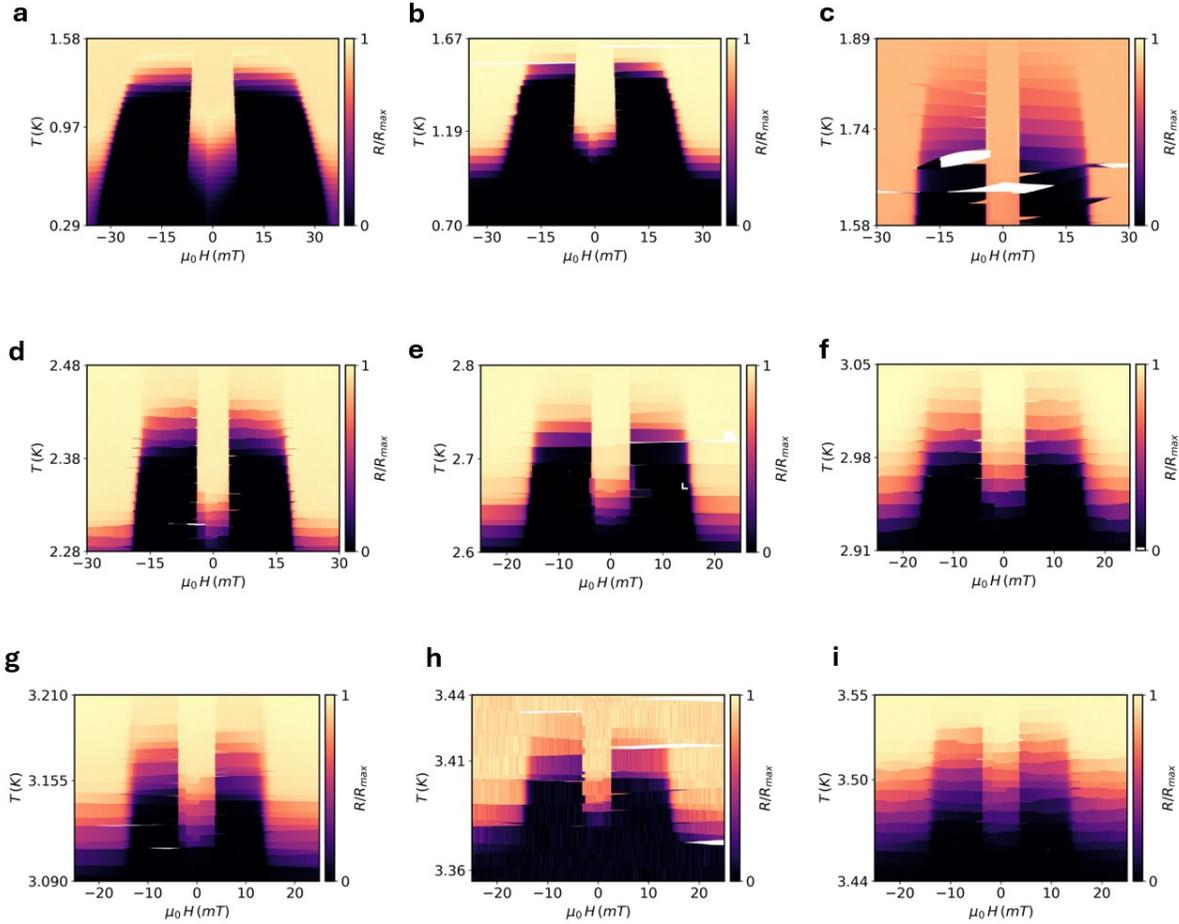

Figure S1: Color map of all other samples of V thickness in nm (9, 9.5, 10, 12, 14, 15, 16, 18 and 20 respectively from a to i. In a) and b) several Resistance (R) vs in plane external magnetic field ($\mu_0 H$) sweeps in the sequence +40mT to -40mT and from -40mT to +40mT (termed downward and upward sweeps respectively) were performed with fixed temperatures; at 5mK intervals for a) from 0.29K to 1.58K, b) from 0.7K to 1.67K. The color map shows the parts of MR measurements corresponding only to 0 to -40mT, and 0 to +40mT, from the downward and upward sweeps, respectively. Similarly, d, e, f, g, h and i represents the color map of other samples. In the colormap, the color represents the Resistance values: the black region indicates the fully superconducting state of the SSV, the purple region represents the transition from the superconducting to the normal state, and the peach color indicates that the superconductor (S) is entirely resistive or in a normal metal state.

Color map of all other V samples whose data are not shown in main paper are presented in Figure 1. Figure S1a and S1b represent the sample with thickness 9 and 9.5 nm, showing a [11]$\Delta T_c$ of about 0.8K and 0.513K, respectively. These samples were measured using the He3 insert which goes up to 0.25K.

Figure S1c presents the Sample with V thickness 10 nm which $T_c$ P is 1.39K and $T_c$ AP is 1.74K because of this we do not have a full color map of this sample as its fall in the temperature range of two different insert one basic which goes to 1.5K and another the He3 insert. From Figure S1d to S1i all these samples were measured in basic insert. There is a clear increase in the $T_c$ value of these samples according to their increase in thickness. From the color maps it is clearly visible that the region where its superconducting is goes on decreasing with increase in the thickness of the V interlayer. In all color map there is clear increase in $T_c$ at the coercive field as the onset of coercive field of 5nm GdN the effective exchange field decreases which gives rise to increase in $T_c$ of the sample. As the magnetic field (H) decreases to 0 mT from +40 mT—where both GdN layers are in a parallel orientation $T_c$ begins to increase due to the reduced exchange field. As the field reaches the coercive point, the value of $T_c$ reaches its maximum, which corresponds to the $T_c$ AP. At this field, the switching of the bulk GdN film occurs, and the two GdN layers enter the antiparallel (AP) state.

## 2. <u>Measurement of MH loops of AlN/3nm GdN/AlN films</u>

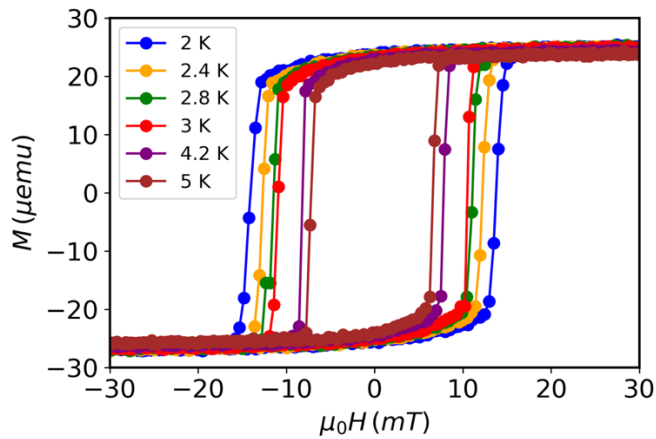

Figure S2. MH loop of 3nm bare GdN layer at different temperatures. The different colour represents the MH loop at that temperature. Coercevity values from these MH loops have been used in Figure 2 of the main manuscript (green filled hexagons).

## 3. <u>Measurement of bare Vanadium films for deriving $T_{c0}$, critical field and dirty limit $\xi_0$</u>

Resistance verses temperature measurement of all V single layers whose data is used for calculating dirty limit coherence length and exchange field are shown in Figure S3.

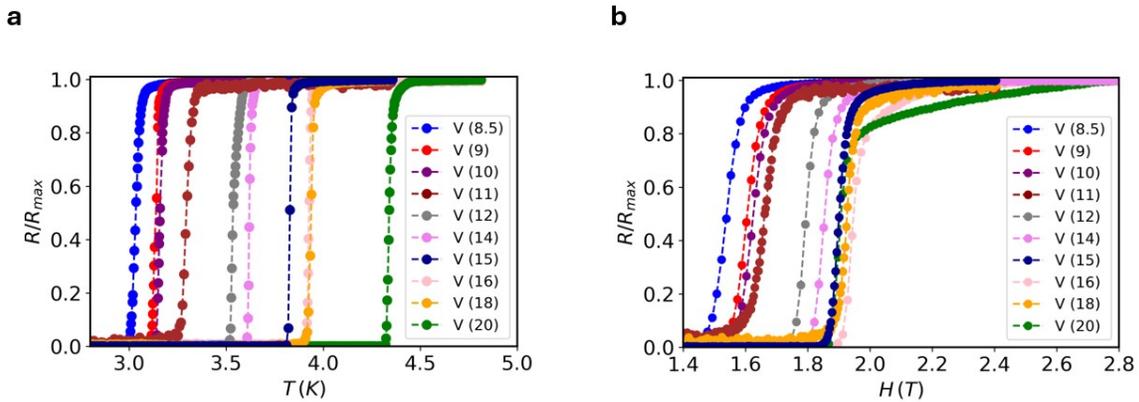

Figure S3.a) Resistance vs temperature heating measurement of all V single layers distinct colour represents different V thickness mentioned in the legend b) Resistance verses applied external magnetic field (out of plane) measurement at temperature 1.6K for all V single layers.

Figure S3a represents the RvsT of all V single layers of thickness (8.5,9,10,11,12,14,15,16,18 and 20) in nm and Figure S3b represents the resistance vs out of plane applied magnetic field. The value of $T_{c0}$ and $H_c$ were used to calculate dirty limit coherence length, exchange field and exchange integral. Table S1 represents the value of $T_{c0}$ and $H_c$(T=1.6K) of V single layers that were used for calculating coherence length for all different thicknesses.

Table S1. Values used for calculating dirty limit $\xi_0$

| Thickness (nm) | $H_{c2}$ (T) (T = 1.6K) | $T_c$ (K) | $H_{c2}$ (T) (T = 0K) | $\xi$ (nm) (T = 0K) |
|---|---|---|---|---|
| 8.5 | 1.54 | 3.03 | 2.731 | 10.978 |
| 9.0 | 1.605 | 3.13 | 2.741 | 10.959 |
| 9.5 | 1.605 | 3.13 | 2.741 | 10.959 |
| 10.0 | 1.63 | 3.158 | 2.756 | 10.928 |
| 11.0 | 1.66 | 3.28 | 2.697 | 11.047 |
| 12.0 | 1.79 | 3.54 | 2.709 | 11.022 |
| 14.0 | 1.85 | 3.62 | 2.748 | 10.943 |
| 15.0 | 1.9 | 3.83 | 2.703 | 11.034 |
| 16.0 | 1.95 | 3.93 | 2.725 | 10.99 |
| 18.0 | 1.93 | 3.94 | 2.692 | 11.056 |
| 20.0 | 1.91 | 4.34 | 2.511 | 11.449 |